\documentclass[reprint,superscriptaddress,amsmath,longbibliography,amssymb,prb]{revtex4-2}

\usepackage{float}
\usepackage{graphicx}
\usepackage{dcolumn}
\usepackage{bm}
\usepackage{multirow}
\usepackage{array}
\usepackage[usenames,dvipsnames]{xcolor}
\definecolor{nblue}{rgb}{0.0, 0.0, 1.0}
\definecolor{magenta}{rgb}{0.79, 0.08, 0.48}
\usepackage[colorlinks,linkcolor=nblue,urlcolor=blue,citecolor=blue,plainpages=false,pdfpagelabels,breaklinks]{hyperref}

\newcommand{\beq}{\begin{equation}}
\newcommand{\eeq}{\end{equation}}
\newcommand{\bea}{\begin{eqnarray}}
\newcommand{\eea}{\end{eqnarray}}

\begin{document}
\title{Superconductivity near 70 K in boron-carbon clathrates MB$_2$C$_8$ (M = Na, K, Rb, Cs) at ambient pressure}

\author{Bin Li}\email[Electronic addresses: ]{libin@njupt.edu.cn}
\affiliation{School of Science, Nanjing University of Posts and Telecommunications, Nanjing 210023, China}
\affiliation{Jiangsu Provincial Engineering Research Center of Low Dimensional Physics and New Energy, Nanjing University of Posts and Telecommunications, Nanjing 210023, China}

\author{Yulan Cheng}
\affiliation{College of Electronic and Optical Engineering, Nanjing University of Posts and Telecommunications, Nanjing 210023, China}

\author{Cong Zhu}
\affiliation{College of Electronic and Optical Engineering, Nanjing University of Posts and Telecommunications, Nanjing 210023, China}

\author{Jie Cheng}\email[Electronic addresses: ]{chengj@njupt.edu.cn}
\affiliation{School of Science, Nanjing University of Posts and Telecommunications, Nanjing 210023, China}
\affiliation{Jiangsu Provincial Engineering Research Center of Low Dimensional Physics and New Energy, Nanjing University of Posts and Telecommunications, Nanjing 210023, China}

\author{Shengli Liu}
\affiliation{School of Science, Nanjing University of Posts and Telecommunications, Nanjing 210023, China}

\date{\today}

\begin{abstract} Inspired by the first boron-carbon (B-C) clathrate SrB$_3$C$_3$ and the ternary borohydride KB$_2$H$_8$ [Miao et al., Phys. Rev. B 104 L100504 (2021)], we have performed first-principles density functional theory calculations of the electronic and phonon band structures for B-C compounds MB$_2$C$_8$ (M = Na, K, Rb, Cs). Our calculations reveal that these materials are dynamically stable and can potentially exhibit superconductivity at ambient pressure. However, only the K, Rb, and Cs compounds exhibit thermodynamic stability below 50 GPa, while NaB$_2$C$_8$ remains thermodynamically unstable at all pressures considered. Based on the Allen and Dynes modified McMillan equation, we predict the superconducting transition temperature $T_c$ of these compounds to be over 65 K at ambient pressure, with $T_c$ decreasing under higher pressures. Remarkably, we find CsB$_2$C$_8$ possesses the highest predicted $T_c$ of 68.76 K. Our findings demonstrate the possibility of high temperature superconductivity in cubic MB$_2$C$_8$ at ambient pressure, expanding the B-C clathrate superconductor family. These results provide valuable insights to guide the identification of new atmospheric pressure superconductors.
\end{abstract}

\maketitle
\section{Introduction}

Ashcroft \emph{et al.} predicted that monatomic hydrogen would transform to the metallic state at high pressure with $T_c$ reaching room temperature \cite{RN135}. The experimental pressure required for the Wigner-Huntington transition to metallic hydrogen may above 495 GPa \cite{RN136,RN137}, which is currently too high to enable practical applications. In recent years, it has been theoretically proposed that metallic hydrogen can be prepared at lower pressure by "chemical precompression" \cite{RN138}, which can be achieved by adding other elements to the hydrogen, allowing the preparation of metallic hydrogen with much lower pressure. Many hydrogen-rich compounds containing main group elements have been considered as potential superconductors with high $T_c$ in the past few years. A large number of binary and ternary hydrides have been theoretically predicted, and significant success has also been achieved experimentally \cite{RN139,RN140,RN141,RN142}. For example, sulfur hydride (H$_3$S) was found be superconducting at 203 K under 155 GPa \cite{RN143,RN144}, and $T_c$ of 250-260 K was observed in lanthanum hydride at 180-170 GPa \cite{RN145,RN146}. However, these binary hydride superconductors need to be formed under extremely high pressure to reach such high transition temperatures. Fortunately, some ternary hydrides, such as LaBH$_8$ \cite{RN147}, LaBH$_9$ \cite{RN148}, LaBeH$_8$ \cite{RN149}, CeBeH$_8$ and CeBH$_8$ \cite{RN150} exhibit stronger "precompression" effects. These compounds are predicted to be potential superconductors and achieve high-temperature superconducting states at relatively low pressures, with some even dynamically stable down to 10 GPa.

However, the majority of the aforementioned hydrogen-rich compounds are formed under specific conditions, and maintaining their stability necessitates a certain pressure. They decompose under low pressures, making it challenging for these systems to exist stably at ambient pressure, which hinders the practical applications \cite{RN151}. Recently, Zhu \emph{et al.} achieved a significant breakthrough by successfully synthesizing a thermodynamically stable carbon-boron clathrate, SrB$_3$C$_3$ \cite{RN152}. Their theoretical predictions suggested that this compound is a promising candidate for phonon-mediated superconductivity at ambient pressure. Subsequently, they embarked on a second round of predictions and synthesis, which confirmed the exceptional properties of SrB$_3$C$_3$ clathrate. Notably, this material exhibits a moderately high superconducting transition temperature ($T_c$) of approximately 22 K at 23 GPa, which is estimated to increase to a remarkable 31 K at ambient pressure \cite{RN153}. In fact, theoretical calculations predict that all of the $Pm\overline{3}n$ XB$_3$C$_3$ alkaline earth analogues (X = Ca, Sr, Ba) exhibit remarkably high $T_c$ in the range of 40-50 K at ambient pressure \cite{RN154,RN155}. Furthermore, the compound Rb$_{0.5}$Sr$_{0.5}$(BC)$_3$ has been predicted to achieve an even higher $T_c$ of approximately 75 K \cite{RN156}.
Compared to their hydride counterparts, boron-carbon superconductors exhibit a significant advantage: they require much lower pressure to achieve superconductivity. Remarkably, some boron-carbon superconductors can even reach a superconducting state at ambient pressure. This is attributed to the unique properties of boron and carbon, which are the lightest elements capable of forming strong covalent bonds. Materials based on these elements are known to be promising candidates for phonon-mediated superconductivity under ambient conditions \cite{RN157}. Therefore, replacing hydrogen with other light elements is considered a feasible way to realize high-$T_c$ superconductivity at ambient pressure.

In this work, we drew inspiration from a previous study ~\cite{RN158}, which predicted that the $T_c$ of KB$_2$H$_8$ compound is 134-146 K at around 12 GPa. To obtain a stable and metallic compound, the authors intercalated a face-centered-cubic (FCC) lattice of potassium with BH$_4$. On this basis, we envisioned replacing hydrogen with the element carbon to obtain new superconductors with lower pressures. We calculated the electronic band structure, phonon spectra and electron-phonon coupling and $T_c$ in MB$_2$C$_8$ (M = Li, Na, K, Rb, Cs) by means of first-principles calculations. Our results indicate that the ambient pressure dynamical stability of MB$_2$C$_8$ simplifies experimental investigations and practical applications by eliminating the need for extreme pressure conditions.

\section{Computational Details}
The full structural optimization, including both the lattice parameters and atomic positions have been calculated within density functional theory as implemented in the QUANTUM ESPRESSO program~\cite{RN159}. Phonon and electron-phonon coupling matrix elements calculations have been carried out using density functional perturbation theory (DFPT)~\cite{RN160}. Pseudopotentials are selected from the Standard Solid State Pseudopotential (SSSP) library~\cite{RN161}. For self-consistent calculations, a $16\times16\times16$ $k$-point grid was used with an energy cutoff of 60 Ry for the wave functions and 480 Ry for the charge density. Dynamical matrices and electron-phonon coupling were calculated on an $8\times8\times8$ $q$-point mesh. A dense $24\times24\times24$ $q$ grid was then used for evaluating an accurate electron-phonon interaction matrix. The effect of electron-phonon coupling on conventional superconductivity can be well represented by the Eliashberg function:
  \begin{equation}
  {\alpha ^2}{\rm{F(}}\omega {\rm{) = }}\frac{1}{{2\pi N(0)}}\sum\limits_{q\nu}^{} {\frac{{\gamma_{q\nu}}}{{\omega_{q\nu}}}\delta (\omega  - \omega_{q\nu})},
  \end{equation}
  Integrating the Eliashberg function over frequency, we can get the electron-phonon coupling  $\lambda$:
  \begin{equation}
  \lambda({\omega}) {\rm{ = }}2\int_0^\infty  {\frac{{{\alpha ^2}{\rm{F}}(\omega )}}{\omega }} d\omega,
  \end{equation}   
To estimate  $T_c$, the Allen-Dynes modified McMillan equation ~\cite{RN166} was used:
\textcolor{black}{
  \begin{equation}
  {T_c} = f_1f_2\frac{{{\omega _{log}}}}{{1.2}}{\rm{exp}}\left[ { - \frac{{1.04(1 + \lambda )}}{{\lambda  - {\mu ^*}(1 + 0.62\lambda )}}} \right],
  \end{equation}
  }
Coulomb pseudopotential $\mu^*$ is taken as 0.1~\cite{RN167}. Electronic structure calculations were performed using the full-potential linearized augmented plane wave (FP-LAPW) method implemented in WIEN2k~\cite{RN162} with the Perdew-Burke-Ernzerhof (PBE) functional~\cite{RN163}. The muffin tin radius was chosen to be 2.3 a.u.\ for M, 1.47 a.u.\ for B and 1.46 a.u.\ for C, respectively. The $R \cdot K_{max}$ = 5.56 , with $R$ is the smallest atomic sphere radius and $K_{max}$ the largest $K$-vector. The crystal structure was visualized using VESTA~\cite{RN164}, while Fermi surfaces were represented with Fermisurfer~\cite{RN165}.

\section{Results and Discussion}
\begin{figure*}[ht]
\begin{center}\includegraphics[width=0.80\textwidth]{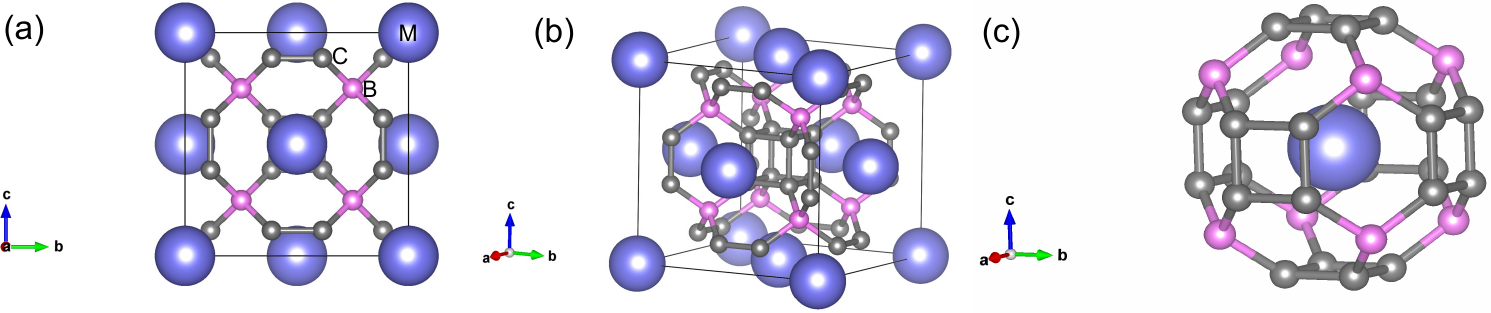}
\caption{\label{fig1} (a) The top view and (b) side view of the crystal structure of MB$_2$C$_8$, where M refers to an alkali metal. The blue, pink and gray spheres represent the M, boron, and carbon atoms, respectively. (c) The B$_{8}$C$_{24}$ cage with M in the center.}
\end{center}
\end{figure*}
\medskip

\vskip 2mm

\begin{table*}[!htbp]
	\centering
	\small
	\caption{\label{table1}Lattice constants and Wyckoff positions of MB$_2$C$_8$ (M = Na, K, Rb, Cs) at ambient pressure and 50 GPa. }
	\begin{tabular}{c c c c c c c }
		\hline
		\hline
		M atoms&Pressure(GPa)&Lattice Parameters (\AA)&Unit-cell volume (\AA$^3$) &Wyckoff positions&Fractional coordinates \\
		\hline
		\multirow{6}*{Na}&\multirow{3}*{0}&\multirow{3}*{a=b=c=6.9808}&\multirow{3}*{340.1854}&B(8c)&-0.25000    0.25000   -0.25000\\
		& & & &C(32f)&-0.38621    0.38621   -0.38621\\
		& & & &Na(4a)&0.00000    0.00000    0.00000\\
		\cline{2-6}
		&\multirow{3}*{50}&\multirow{3}*{a=b=c=6.6109}&\multirow{3}*{288.9228}&B(8c)&-0.25000    0.25000   -0.25000\\
		& & & &C(32f)&-0.38438    0.38438   -0.38438\\
		& & & &Na(4a)&0.00000    0.00000    0.00000\\
		\hline
		\multirow{6}*{K}&\multirow{3}*{0}&\multirow{3}*{a=b=c=7.0237}&\multirow{3}*{346.4957}&B(8c)&0.25000   -0.25000    0.25000\\
		& & & &C(32f)&0.38654   -0.38654    0.38654\\
		& & & &K(4a)&0.00000    0.00000    0.00000\\
		\cline{2-6}
		&\multirow{3}*{50}&\multirow{3}*{a=b=c=6.6508}&\multirow{3}*{294.1858}&B(8c)&0.25000    -0.25000   0.25000\\
		& & & &C(32f)&-0.38465    0.38465   -0.38465\\
		& & & &K(4a)&0.00000    0.00000    0.00000\\
		\hline
		\multirow{6}*{Rb}&\multirow{3}*{0}&\multirow{3}*{a=b=c=7.0640}&\multirow{3}*{352.4943}&B(8c)&-0.25000    0.25000    0.25000\\
		& & & &C(32f)&-0.38685    0.38685    0.3868\\
		& & & &Rb(4a)&0.00000    0.00000    0.00000\\
		\cline{2-6}
		&\multirow{3}*{50}&\multirow{3}*{a=b=c=6.6894}&\multirow{3}*{299.3378}&B(8c)&-0.25000    0.25000   0.25000\\
		& & & &C(32f)&-0.38494    0.38494   -0.38494\\
		& & & &Rb(4a)&0.00000    0.00000    0.00000\\
		\hline
		\multirow{6}*{Cs}&\multirow{3}*{0}&\multirow{3}*{a=b=c=7.1266}&\multirow{3}*{361.9488}&B(8c)&-0.25000    0.25000    0.25000\\
		& & & &C(32f)&-0.38736    0.38736    0.3873\\
		& & & &Cs(4a)&0.00000    0.00000    0.00000\\
		\cline{2-6}
		&\multirow{3}*{50}&\multirow{3}*{a=b=c=6.7467}&\multirow{3}*{307.0960}&B(8c)&-0.25000    0.25000   0.25000\\
		& & & &C(32f)&-0.38536    0.38536   -0.38536\\
		& & & &Cs(4a)&0.00000    0.00000    0.00000\\
		\hline
	    \hline
	\end{tabular}
	\end{table*}

	\medskip
\medskip

\vskip 2mm


\begin{table}[!htbp]
	\caption{\label{table2}Atomic distances of B-C and C-C in MB$_2$C$_8$ at ambient pressure and 50 GPa.}
	\centering
	
	\begin{tabular}{c c c c   }
		\hline
		\hline
		 &Pressure (GPa)&d$_{B-C}$ (\AA)&d$_{C-C}$ (\AA)\\
		\hline
		\multirow{2}*{Na}&0&1.6469&1.5887\\
		\cline{2-4}
		&50&1.5387&1.5287\\
		\hline
		\multirow{2}*{K}&0&1.6611&1.5938\\
		\cline{2-4}
		&50&1.5511&1.5343\\
		\hline
		\multirow{2}*{Rb}&0&1.6744&1.5986\\
		\cline{2-4}
		&50&1.5635&1.5394\\
		\hline
		\multirow{2}*{Cs}&0&1.6955&1.6055\\
		\cline{2-4}
		&50&1.5818&1.5469\\
	
		\hline
		\hline
	\end{tabular}
\end{table}

\medskip
\medskip

\begin{figure*}[ht]
\begin{center}\includegraphics[width=0.80\textwidth]{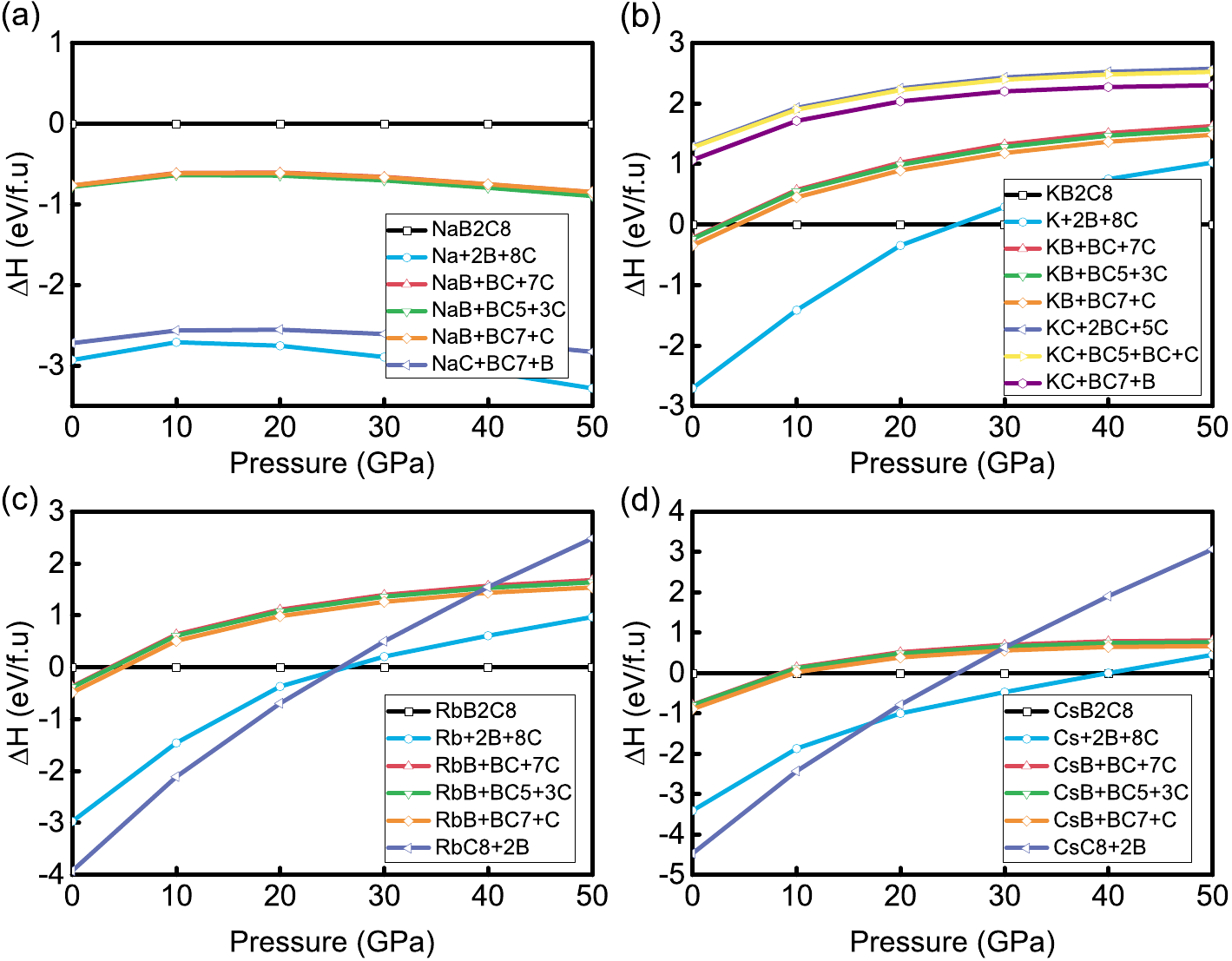}
\caption{\label{fig2}Calculated enthalpy change ($\Delta H$) as a function of pressure for MB$_2$C$_8$ compounds with M = Na, K, Rb, Cs. $\Delta H$ is referenced relative to the enthalpy of decomposition into standard elemental phases decompositions at the corresponding pressure. Other possible decomposition enthalpies were also considered. More negative $\Delta H$ indicates greater thermodynamic stability. Figures (a)-(d) show $\Delta H$ plots for NaB$_{2}$C$_{8}$, KB$_{2}$C$_{8}$, RbB$_{2}$C$_{8}$, and CsB$_{2}$C$_{8}$ respectively}
\end{center}
\end{figure*}

Figure \ref{fig1} illustrates the crystal structures of MB$_2$C$_8$ crystallizing in the cubic $Fm\overline{3}m$ space group (No. 225). The M, B and C atoms occupy the 4$a$ (0.00, 0.00, 0.00), 8$c$ (-0.25, 0.25, 0.25), and 32$f$ ($-0.38\pm u,0.38\pm u,0.38\pm u$) Wyckoff positions, respectively, where the variable $u$ depends on pressure and the metal species M (see Table \ref{table1}). The boron and carbon atoms form a B$_{8}$C$_{24}$ cage-like arrangement surrounding the metal atoms. We compute the formation enthalpies of the ternary carbides relative to decomposition into binary precursors, taking crystal structures of the binaries from the Materials Project database\cite{Jain2013}: $Pm\overline{3}m$ (M, MB), $Immm$ (NaC), $Pm\overline{3}m$ (KC), $Pnma$ (RbC, CsC), $P6/mmm$ (RbC$_8$, CsC$_8$), $P6{3}mc$ (BC), $P3m1$ (BC$_5$), and $P\overline{4}m2$ (BC$_7$) within 0-50 GPa. Pressure-dependent enthalpy differences are plotted in Fig. \ref{fig2}, showing that NaB$_2$C$_8$ is thermodynamically unstable up to 50 GPa while KB$_2$C$_8$, RbB$_2$C$_8$, and CsB$_2$C$_8$ decompose above 25, 25, and 40 GPa, respectively. Ambient and high-pressure lattice constants presented in Table \ref{table1} show unit cell volumes expand with increasing atomic number $Z$, while lattice parameters concurrently exhibit compression under pressure. Interatomic distances in Table \ref{table2}) demonstrate a corresponding reduction under applied pressure and enhancement as a function of increasing $Z$.

\vskip 4mm

\medskip

\begin{figure*}[ht]
	\begin{center}\includegraphics[width=0.80\textwidth]{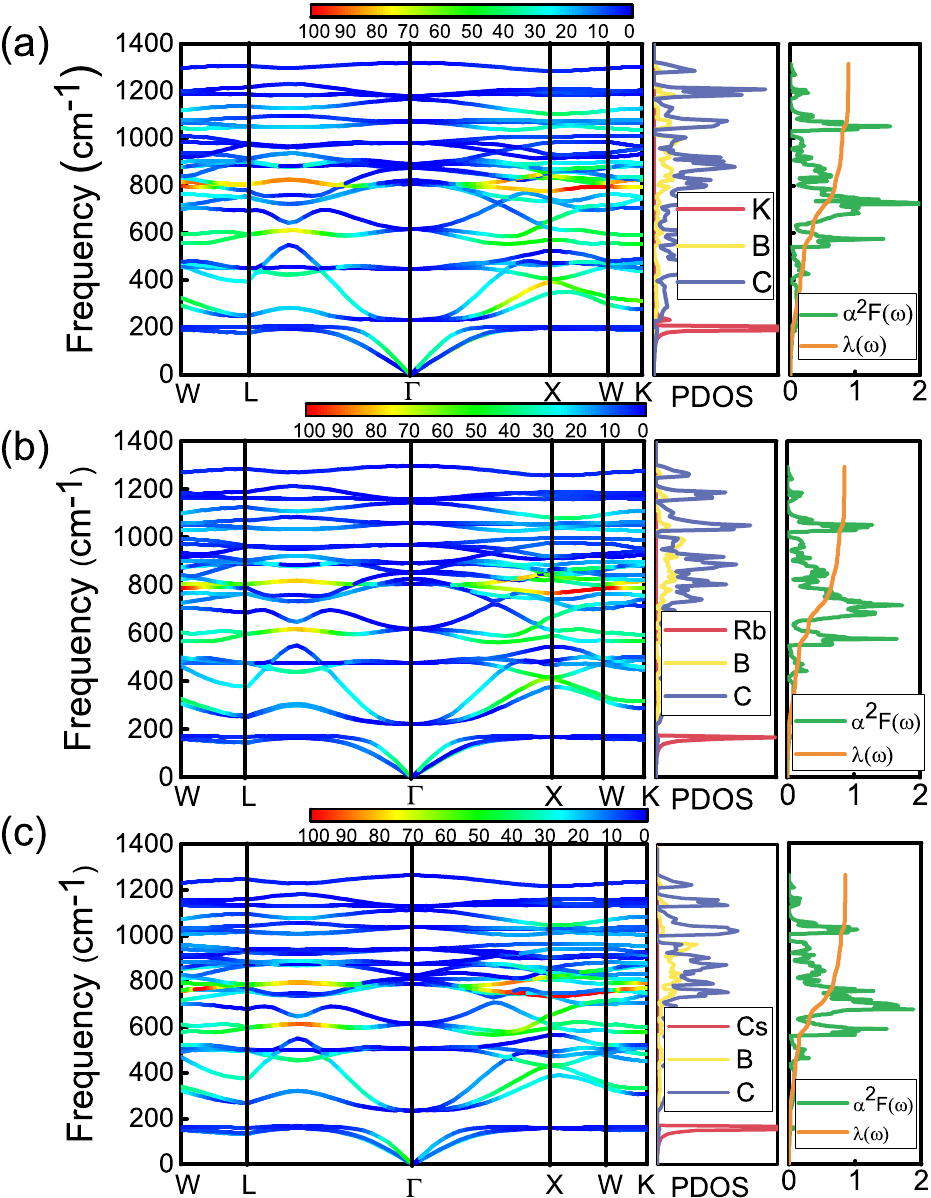}
		\caption{\label{fig3}(a-c) Phonon dispersion curves showing the phonon frequencies ($\omega_{q\nu}$) with a color scale representing the phonon linewidths ($\lambda_{q\nu}\omega_{q\nu}$), the phonon density of states (PHDOS) projected onto individual atoms, the Eliashberg spectral function $\alpha^2F(\omega)$, and the electron-phonon coupling strength $\lambda(\omega)$ at 40 GPa for (a) KB$_2$C$_8$, (b) RbB$_2$C$_8$ and (c) CsB$_2$C$_8$. In the PHDOS plots, the projections onto the K, Rb, Cs (red), B (yellow), and C (blue) atoms are shown in different colors to distinguish the contributions of different atoms.}
	\end{center}
\end{figure*}

The thermodynamic stability of MB$_2$C$_8$ above 40 GPa motivate the examination of phononic properties. Phonon dispersions of $Fm\overline{3}m$-MB$_2$C$_8$ (M = Li, Na, K, Rb, Cs) were calculated using the DFPT methodology at pressures of 0-50 GPa. The calculation results  indicate stable phonon dispersion across the studied pressure regime with the sole exception of LiB$_2$C$_8$ (see Supplementary Figure S1 for Li data)\cite{som}. The phonon dispersions and density of states at 40 GPa (Fig. \ref{fig3}) reveal dynamical stability under compression, evidenced by the absence of imaginary phonon frequencies along high-symmetry $q$ path. The phonon spectra segregate into two regimes: (i) Low-frequency acoustic branches below 200 cm$^{-1}$ stemming predominantly from alkali metal (M) atom oscillations, with heavier K, Rb, and Cs atoms inducing lower frequency modes. This manifests as the sharp low frequency peak in the density of states (DOS) plots. (ii) High-frequency optical branches between 500-1400 cm$^{-1}$ associated with vibrations of the covalently bonded borocarbon cages.
The alkali atom vibrations contribute primarily to acoustic modes and low-frequency density of states peaks, while the borocarbon cage vibrations dominate the higher frequency optical phonon branches. The electron-phonon coupling enhancement from these high frequency B and C vibrations further influences emergent properties of select MB$_2$C$_8$ compounds under compression. 

Analysis of the $\Gamma$-point optical branches reveals a declining maximum phonon frequency when progressing from lighter to heavier alkali metals in the MB$_2$C$_8$ compounds. The highest $\Gamma$-point optical frequency decreases from 1319.024 cm$^{-1}$ in KB$_2$C$_8$ down to 1263.786 cm$^{-1}$ for CsB$_2$C$_8$. Meanwhile, a small gap emerges in the phonon spectra, separating low-frequency acoustic branches below 200 cm$^{-1}$ from the higher optical branches. This gap widens as the alkali metal atomic number increases, spanning just 13.37 cm$^{-1}$ in KB$_2$C$_8$ but reaching 59.76 cm$^{-1}$ in CsB$_2$C$_8$. \textcolor{black}{As the atomic mass of the central metal increases, the lattice vibrations weaken, leading to a decrease in frequency and softening of low-frequency phonons. The widening of the gap is also attributable to the slight softening of low-frequency acoustic phonons. However, the effect of phonon softening on superconductivity varies under different pressure conditions. At ambient pressure, the superconducting transition temperature rises with increasing atomic mass, but it decreases at pressures around 50 GPa. This contrasting behavior may be related to the varying pressure dependence of the superconducting transition temperature for different central atoms. }

\medskip

The phonon density of states curves demonstrate that carbon atoms contribute more strongly than boron atoms. As shown in Figure \ref{fig3}, the alkali metal atoms do not significantly impact the electron-phonon coupling $\lambda$. In KB$_2$C$_8$ compound, vibrations of the B-C network account for approximately 87\% of $\lambda$, while K atoms contribute only ~13\%. This suggests superconductivity arises predominantly from vibrations of the B-C sublattice. Phonon dispersion calculations reveal strong electron-phonon coupling between the $X$ and $W$ points in the Brillouin zone near 800 cm$^{-1}$, \textcolor{black}{and the $W$-$L$ and $L$-$\Gamma$ directions also exhibit considerable coupling strength. The $L$-$\Gamma$ direction in the Brillouin zone around 600 cm$^{-1}$ exhibits a relatively strong electron-phonon coupling strength, which is further enhanced  in CsB$_2$C$_8$.} The Eliashberg function, $\alpha^2F(\omega)$, contains three principal peaks that can be categorized into three frequency regions: [400-600 cm$^{-1}$], [600-1000 cm$^{-1}$], and [1000-1400 cm$^{-1}$]. Below 400 cm$^{-1}$, the electron-phonon coupling is weak, consistent with the minor contributions from alkali metal vibrations. In contrast, the predominant electron-phonon coupling intensity resides above the phonon frequency gap, arising primarily from vibrations of the B-C vibrations.

\begin{figure*}[ht]
	\begin{center}\includegraphics[width=0.6\textwidth]{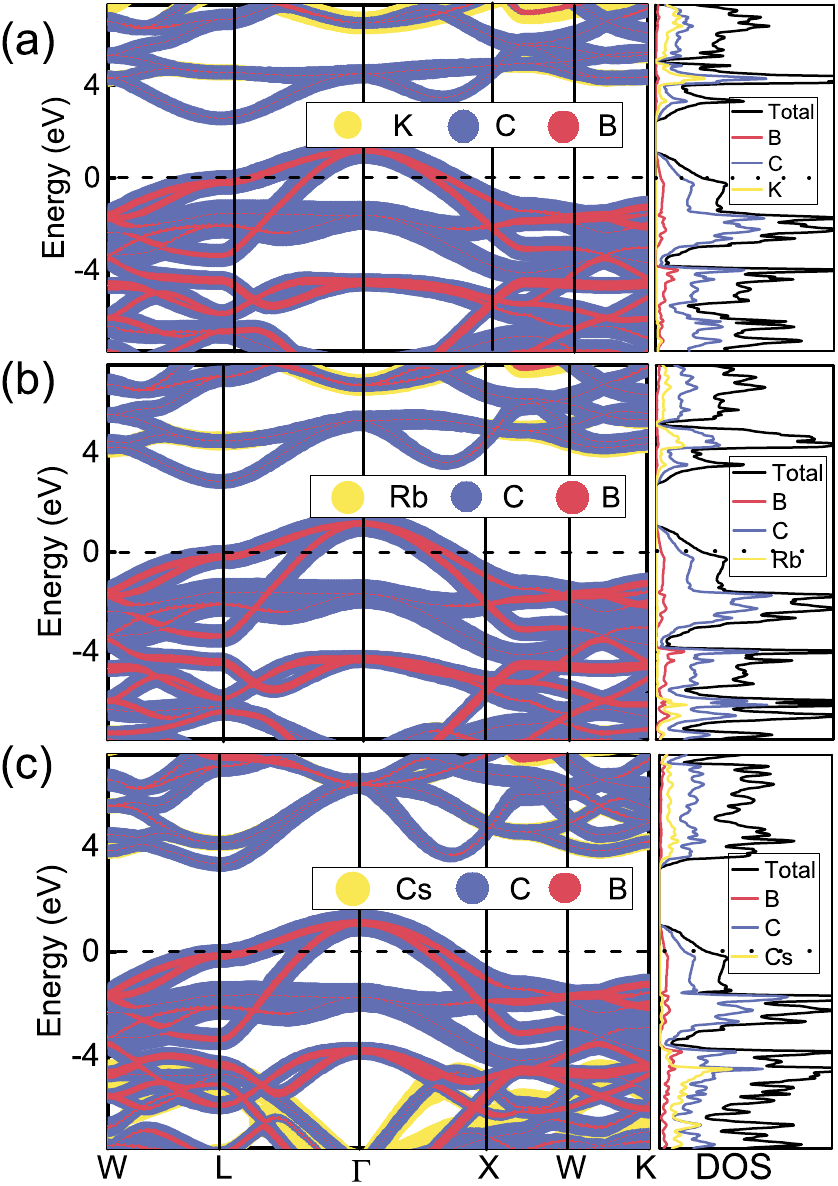}
		\caption{\label{fig4}Electronic band structures and partial density of states (DOS) projected on individual atomic species for MB$_2$C$_8$ (M = K, Rb, Cs) at 40 GPa pressure for KB$_2$C$_8$, (b) RbB$_2$C$_8$, and (c) CsB$_2$C$_8$. In the DOS plots, the projections onto the M (K/Rb/Cs), B, and C atoms are shown in yellow, blue and red colors respectively. The Fermi energy is set to zero on the energy scale.}
	\end{center}
\end{figure*}	 
\medskip

\vskip 4mm

\begin{figure}[ht]
	\begin{center}\includegraphics[width=0.50\textwidth]{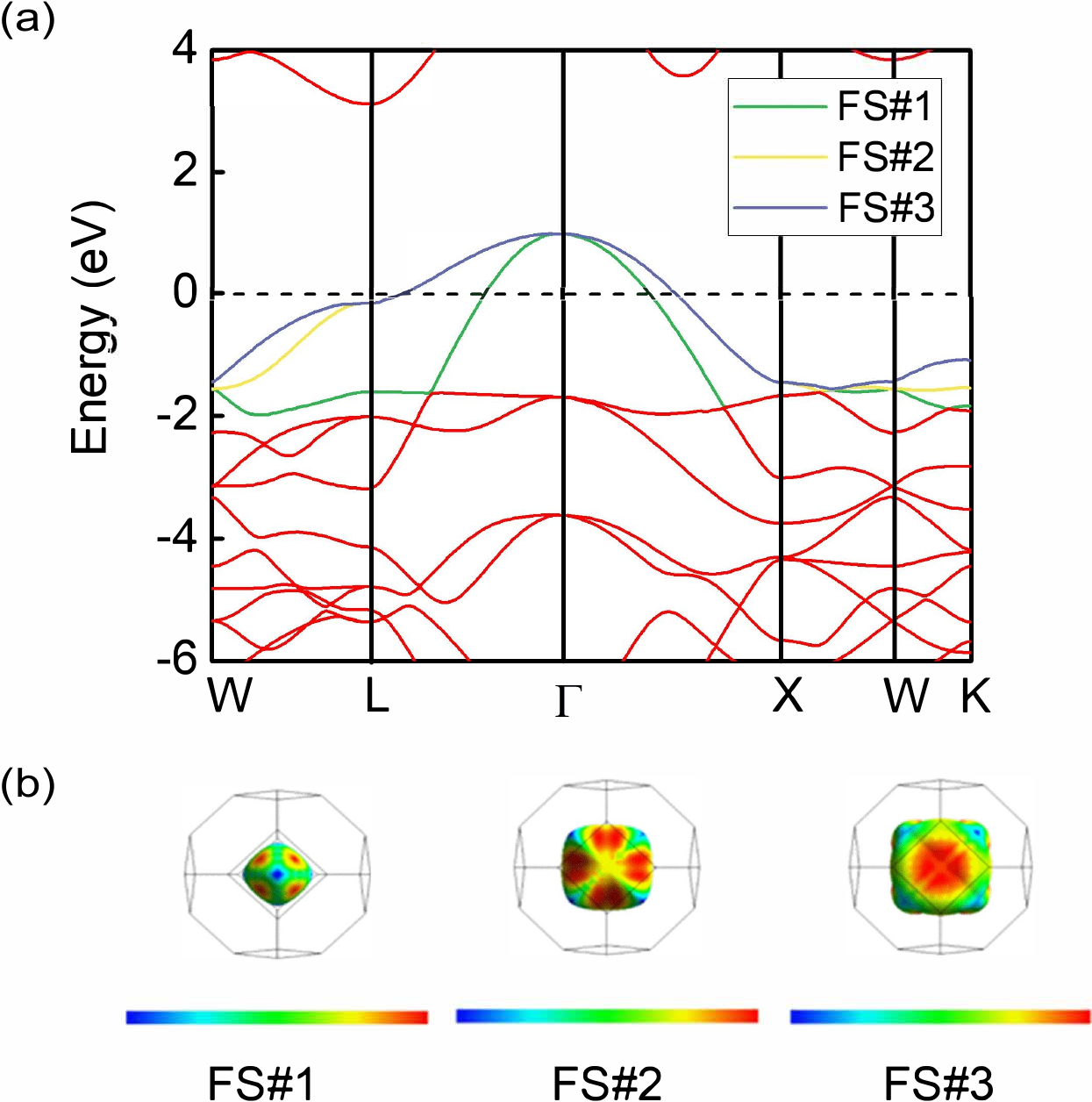}
		\caption{\label{fig5}(a) Electronic band structures of CsB$_2$C$_8$ at 40 GPa. (b) The three Fermi surfaces (FS\#1, FS\#2, FS\#3) derived from bands crossing the Fermi level in (a). The FS contours are colored according to the Fermi velocities.}
	\end{center}
\end{figure}
\medskip

We calculated the electronic band structures and partial density of states projected onto individual atoms for MB$_2$C$_8$ (M = K, Rb, Cs) under pressure, shown in figure \ref{fig4}. Despite having different alkali metals M and external pressures, their band structures share similar characteristics, exhibiting metallic behavior with three bands crossing the Fermi level along the high-symmetry $k$ path. In the region near the Fermi level, the DOS originates predominantly from the B and C atoms, with a greater contribution coming from the more electronegative C atoms~\cite{RN155}. Additionally, as the alkali metal changes from K to Rb to Cs, the indirect pseudo bandgap above the Fermi level increases progressively from 1.38 to 2.23 eV.

Figure \ref{fig5} shows the band structure and corresponding Fermi surfaces (FSs) of CsB$_2$C$_8$. The green, blue and yellow colored bands in Figure \ref{fig5}(a) correspond to the three FSs shown in Figure \ref{fig5}(b). FS\#1 has a dice-like morphology, with the minimum Fermi velocity at the vertex and maximum velocity at the center of each triangular face. FS\#2 forms a concave cubic shape, with velocity maxima located at the inward-concave regions of each cube face. Lastly, FS\#3 resembles a cubic morphology as well, but with both vertex and face-center regions concavely indented. For FS\#3, the minimum velocities occur at the concave vertices while maximum velocities are found at the cube face centers.
 \vskip 2mm


\begin{table*}[!htbp]
	\caption{\label{table3}The calculated superconducting critical temperature $T_c$ of MB$_2$C$_8$ (M = Na, K, Rb, Cs) at various pressures.}
	\centering
	
	\begin{tabular}{c c c c c c c  }
		
		\hline
		\hline
		
		\multirow{2}*{  }&\multicolumn{6}{c}{$T_c$(K)}  \\
		\cline{2-7}
		&0 (GPa)&10 (GPa)&20 (GPa)&30 (GPa)&40 (GPa)&50 (GPa)\\
		\hline
		Na&65.30&61.13&57.69&55.00&52.66&50.68\\
		K&65.55&60.92&56.90&53.52&50.69&48.36\\
		Rb&66.56&61.47&56.86&52.78&49.61&46.85\\
		Cs&68.76&62.59&57.91&52.79&49.67&46.16\\
		
		\hline
		\hline
		
	\end{tabular}
\end{table*}

\begin{figure}[ht]
\begin{center}\includegraphics[width=0.50\textwidth]{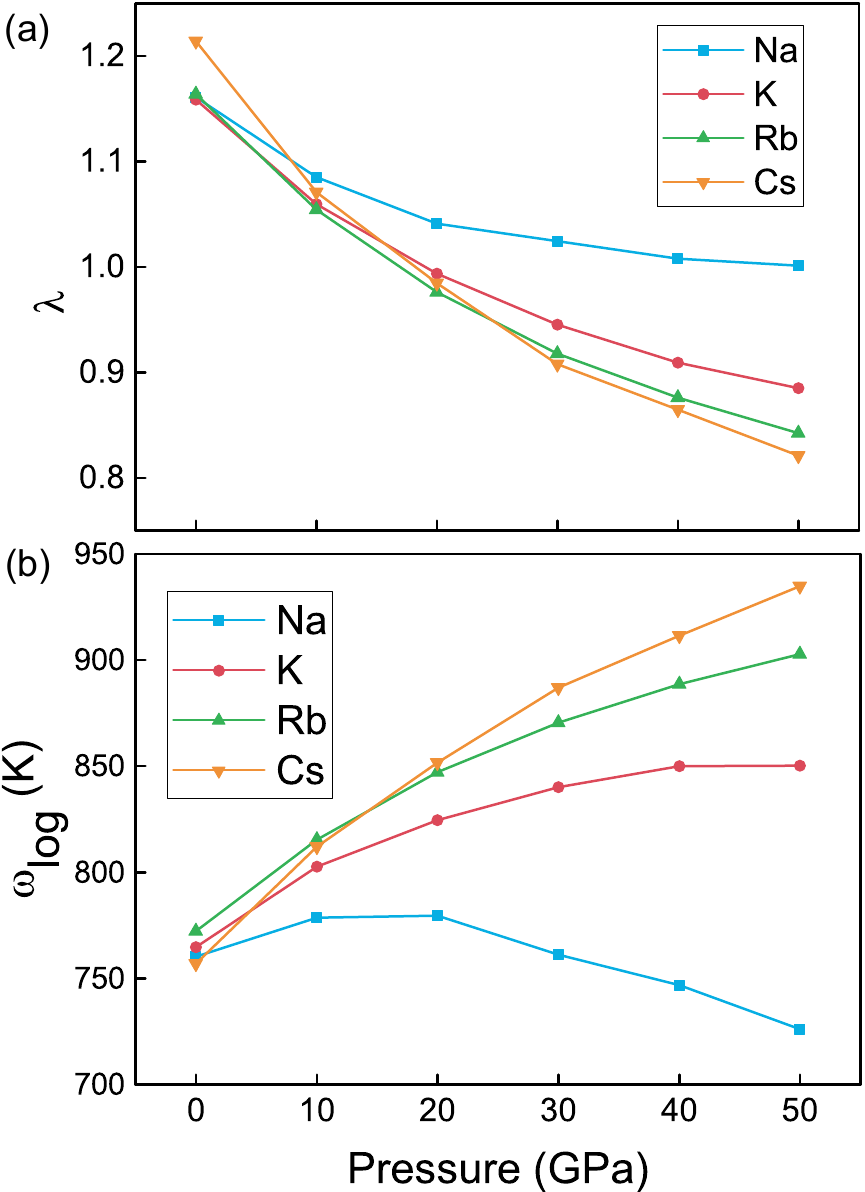}
 \caption{\label{fig6}(a) Calculated electron-phonon coupling constant $\lambda$, and (b) logarithmic average of phonon frequency $\omega_{log}$ for MB$_2$C$_8$ (M = Na, K, Rb, Cs) as a function of pressure.}
 \end{center}
\end{figure}

\medskip
\medskip 
The superconducting transition temperatures of MB$_2$C$_8$ at 0-50 GPa are listed in Table \ref{table3}, which exhibit a maximum predicted $T_c$ of 68.76 K at 0 GPa in CsB$_2$C$_8$. As external pressure rises from 0 to 50 GPa, $T_c$ exhibits  a declining trend. \textcolor{black}{We also employed the original McMillan equation to estimate the superconducting transition temperature, with the results presented in the supplementary material. The calculated superconducting transition temperature is slightly lower compared to the values reported in Table \ref{table3}.} Figure \ref{fig6}(a) shows the electron-phonon coupling $\lambda$ mirrors this pressure dependence, diminishing progressively across the same pressure range. However, as shown in Figure \ref{fig6}(b), the logarithmic average phonon frequency $\omega_{log}$ shows the opposite behavior, increasing with mounting pressure for the most cases. The one exception is NaB$_2$C$_8$, for which $\omega_{log}$ declines after 20 GPa. At ambient pressure, substituting heavier alkali metals causes a slight enhancement in $\lambda$ and $T_c$. However, contrasting different compounds at pressures exceeding 20 GPa reveals that substituting heavier alkali metals causes a slight reduction in $\lambda$ and $T_c$, while $\omega_{log}$ increases progressively down the alkali metal series from Na to Cs. Given the consistent correlation observed between the pressure and alkali metal dependence of both $T_c$ and $\lambda$, it can be inferred that the electron-phonon coupling strength plays the predominant role in influencing the superconducting transition temperature.

 \medskip

In summary, we employed first-principles methods to investigate the fundamental properties of the alkali metal-boron-carbon compound MB$_2$C$_8$ (M = Li, Na, K, Rb, Cs) across a pressure range of 0-50 GPa. Our findings indicate that all compounds, except for LiB$_2$C$_8$, reside in a metastable state under ambient conditions. Notably, MB$_2$C$_8$ (M = Na, K, Rb, Cs) exhibits dynamic stability within this pressure range, while NaB$_2$C$_8$ is thermodynamically unstable. KB$_2$C$_8$ and RbB$_2$C$_8$ become thermodynamically stable at pressures exceeding 25 GPa, while CsB$_2$C$_8$ forms with a positive enthalpy at 40 GPa, suggesting their potential synthesis at high pressures followed by quenching to ambient conditions. Electron-phonon calculations which mainly from vibrations of borocarbon cage, indicate  that MB$_2$C$_8$ compounds could exhibit superconductivity with a maximum critical temperature of 68.76 K at ambient pressure. Our prediction of these novel  MB$_2$C$_8$ compounds opens up further possibilities for the application of ternary B-C clathrates.
\newpage

\begin{acknowledgments}
This work is supported by the National Natural Science Foundation of China (Grants No. 12175107) and Nanjing University of Posts and Telecommunications Foundation (NUPTSF) (Grant No. NY219087, NY220038). Some of the calculations were performed on the supercomputer in the Big Data Computing Center (BDCC) of Southeast University.
\end{acknowledgments}
%

\end{document}